\documentclass[a4paper,11pt,aps,preprint]{revtex4}
\usepackage{lineno,hyperref}

\usepackage{graphicx,times,nicefrac,amsmath,amsfonts,amssymb,amsthm,epsf,bm,bbm,longtable,dcolumn}
\usepackage{revsymb,wasysym,mathrsfs,epsf,color}
\usepackage{graphicx,color,graphics,epsf,dcolumn}
\DeclareGraphicsExtensions{.pdf}

\begin{document}
\title{Main magnetic focus ion source: Device with high electron current density}

\author{V.P.~Ovsyannikov}\thanks{URL: \url{http://mamfis.net/ovsyannikov.html}}
\affiliation{MaMFIS Group, D-01069  Dresden, Germany \\
Joint Institute for Nuclear Research, 141980  Dubna, Russia}

\author{A.V.~Nefiodov}\thanks{Corresponding author. E-mail: anef@thd.pnpi.spb.ru}
\affiliation{Petersburg Nuclear Physics Institute, 188300 Gatchina, St.~Petersburg, Russia}

\author{A.Yu.~Boytsov}
\affiliation{Joint Institute for Nuclear Research, 141980  Dubna, Russia}
\author{A.Yu.~Ramzdorf}
\affiliation{Joint Institute for Nuclear Research, 141980  Dubna, Russia}
\author{V.I.~Stegailov}
\affiliation{Joint Institute for Nuclear Research, 141980  Dubna, Russia}
\author{S.I.~Tyutyunnikov}
\affiliation{Joint Institute for Nuclear Research, 141980  Dubna, Russia}

\author{A.A.~Levin}
\affiliation{Ioffe Institute, 194021 St.~Petersburg, Russia}

\widetext


\begin{abstract}
We discuss recent experiments performed with an upgraded  version of the main magnetic focus ion source (MaMFIS) at the Joint Institute for Nuclear Research (JINR) in Dubna. The device operates in the range of electron beam energies extended up to $40$~keV.  The achieved electron current densities are of the order of $10$ kA~cm$^{-2}$. This assessment is consistent  both with the very short ionization time and with the utmost ionization degree of the produced highly charged ions.  Due to its high efficiency, the  MaMFIS technology is especially promising for ionization of  short-lived radionuclides and heavy elements. A new scheme for charge breeding is proposed. 
\end{abstract}

\maketitle

\section{Introduction}

At present, highly charged ions are the subject of extensive research and numerous applications in atomic, nuclear, and particle physics \cite{wenander10,rodriguez10,simon12,thorn12,boytsov15,lapierre18,kozlov18,lapierre19}.  High-quality beams of ions with different ionization degree for any elements and isotopes, stable and radioactive, are required.  The low-energy highly charged ions are produced in ion sources both from neutral atomic targets and from externally injected low (more often singly) charged ions. In the latter case, the charge-state transformation is known as charge breeding. The most-widely-used ion sources (electron beam and electron-cyclotron resonance) operate on the basis of the successive ionization by electron impact.

A particular choice of specific characteristics of the ion beam and corresponding conditions of its production in ion sources can bring  significant advantage. For example, higher ion charges $q$ allow one to increase the precision of mass measurements in  Penning traps \cite{blaum20}. The beams of radioactive nuclides should be bred with high efficiency much faster than they decay. For heavy isotopes with short half-lives, multiple charge stripping to $q \gg +1$ within the confinement time of  about 1~ms still poses a serious challenge  as it requires a high current density $j_e$ of the electron beam.

\begin{figure}[t!]    
\begin{minipage}[t]{\textwidth}
\centering\includegraphics[width=0.44\columnwidth,angle=0,keepaspectratio,trim=80  15  430  20,clip]{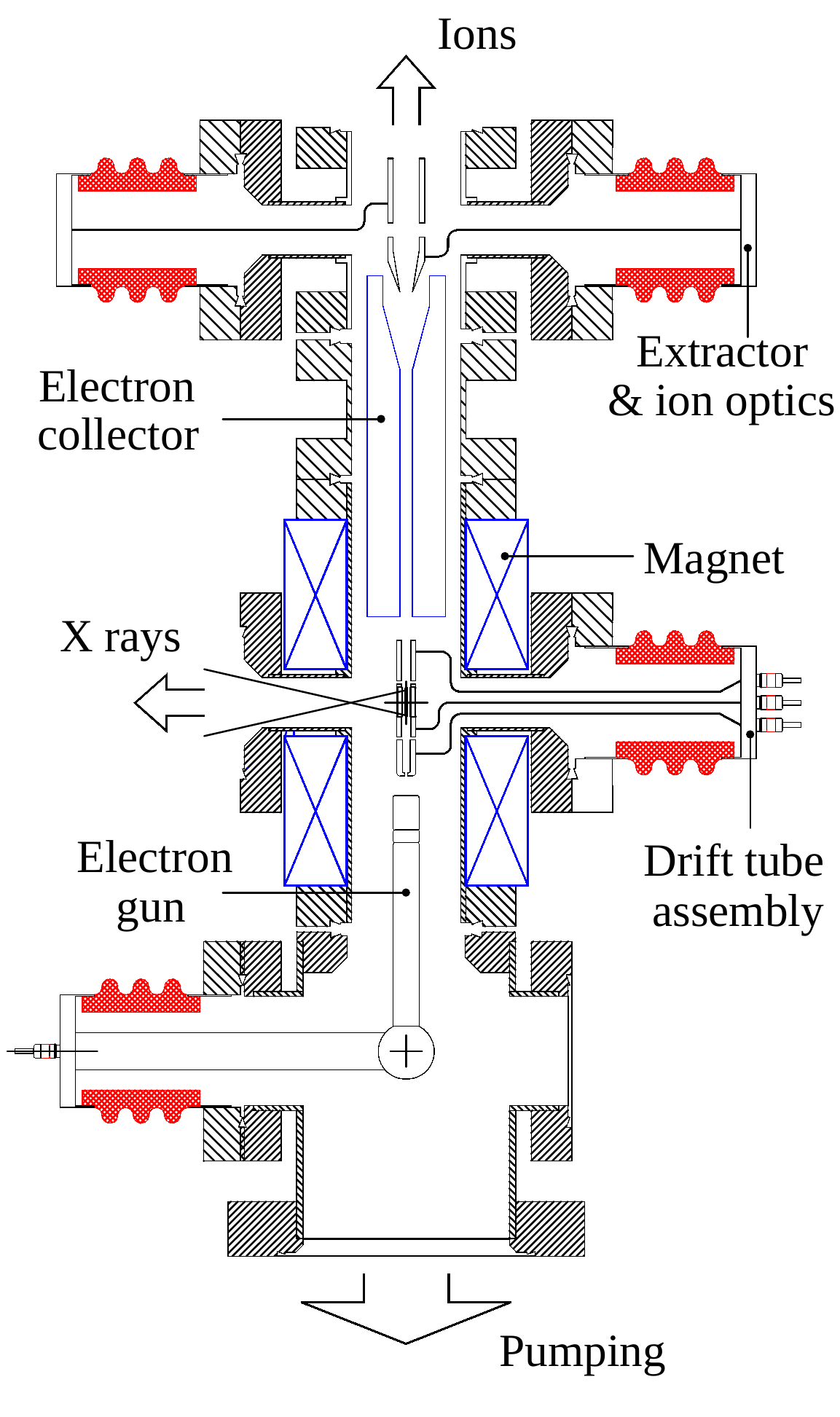}
\end{minipage}
\begin{minipage}[t]{\textwidth}
\centering\includegraphics[width=0.44\columnwidth,angle=0,keepaspectratio,clip]{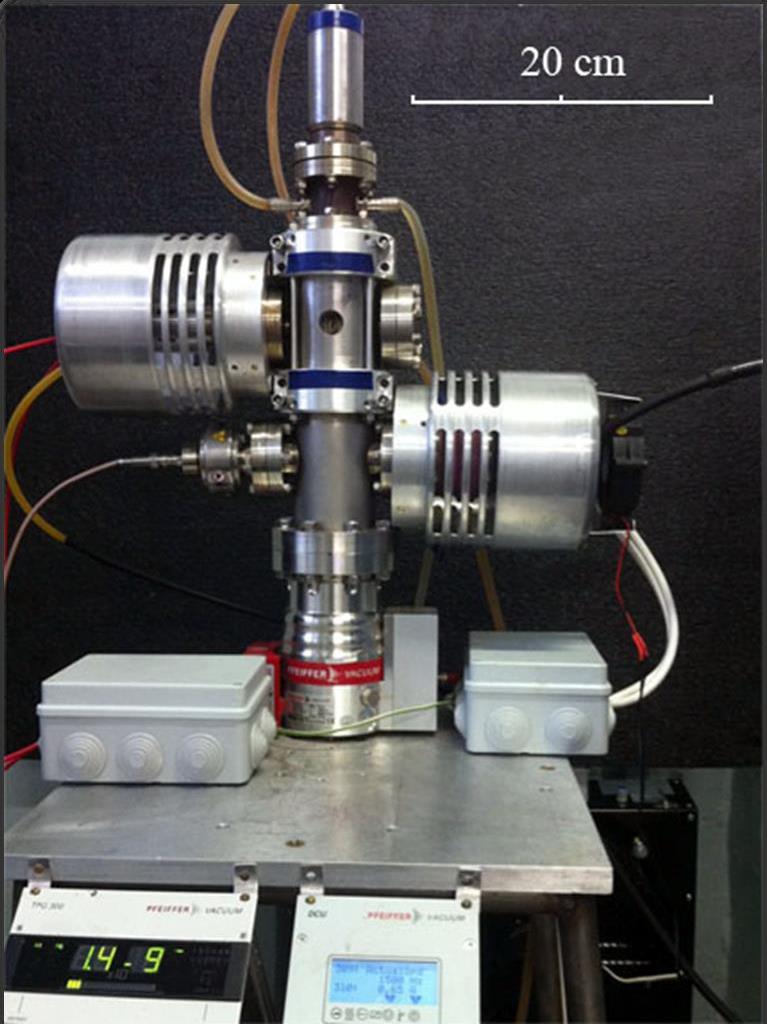}
\end{minipage}
\caption{\label{fig1} Dubna MaMFIS: principle scheme and general view (photo) of the device.}
\end{figure}

In this work, we describe a further development of the main magnetic focus ion source (MaMFIS), a compact room-temperature  device characterized by the operating parameter $j_e \sim 10$~kA~cm$^{-2}$. Such high electron current densities  are achieved in local ion traps  formed in crossovers of the rippled electron beam focused by thick magnetic lens  \cite{ovsyannikov16a,ovsyannikov16b}. This should be compared with the electron-beam ion source (EBIS) technology, which employs smooth electron flows and can attain the highest electron current density in the case of the Brillouin focusing  only  \cite{donets67,donets98}. In a crossover, the Brillouin limit is not justified and the electron current density can exceed it by orders of magnitude. We assess the magnitude of $j_e$  indirectly by means of two quantities, namely, the ionization time and the utmost achievable charge state of the produced highly charged ions.  The degree of ionization is determined from the high-energy part of  the X-ray spectrum corresponding to the radiative recombination of highly charged ions with beam electrons. We suggest a novel scheme for charge breeding, which is most promising to use in the rare-isotope beam facilities, which require low-intensity bunches of highly charged ions to be produced at a high repetition rate.

\begin{figure}[t!]    
\begin{minipage}[t]{\textwidth}
\centering\includegraphics[width=0.99\columnwidth,angle=0,keepaspectratio,trim=11  100  0  100,clip]{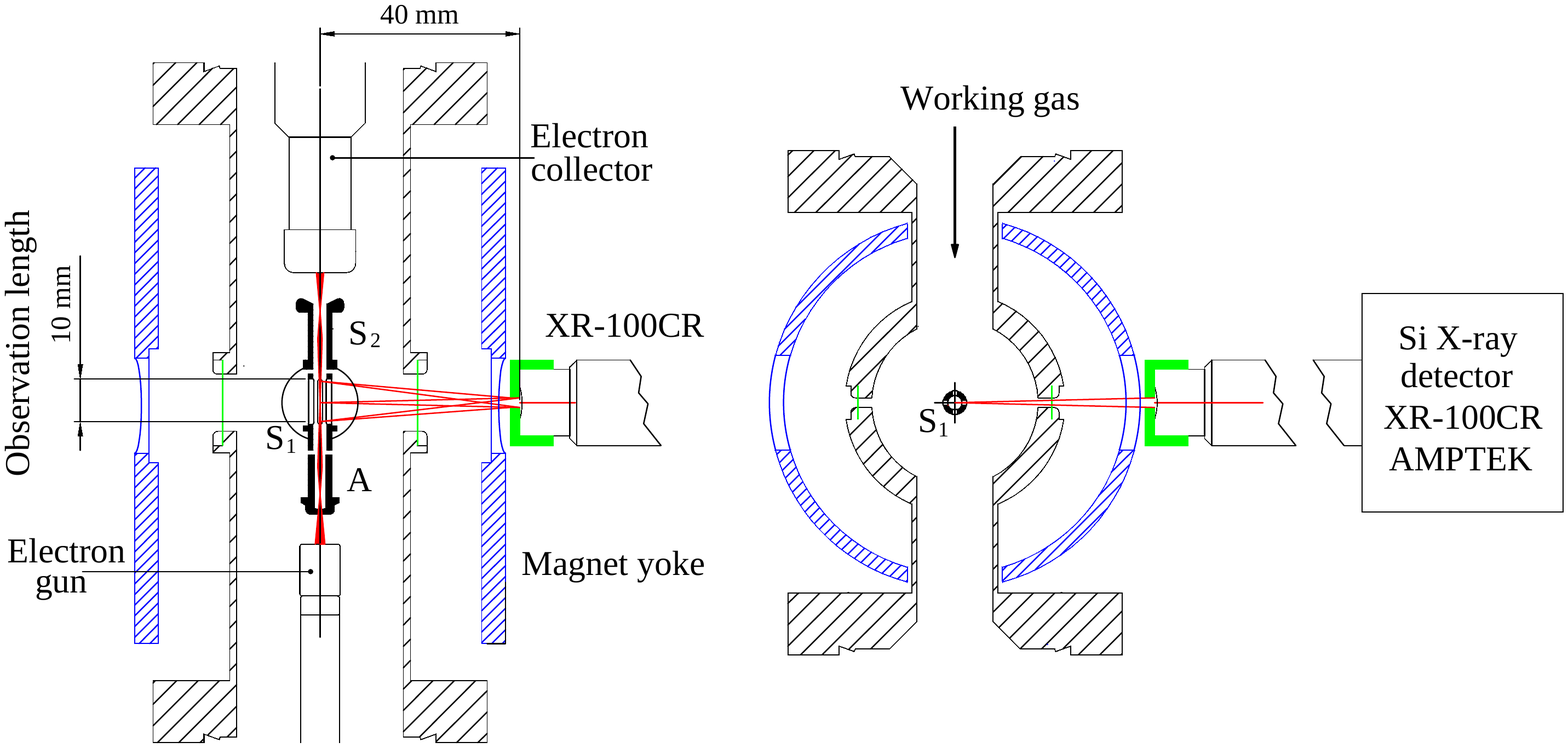}
\end{minipage}
\caption{\label{fig2}  General scheme of experiment (views from the side and from the top). Anode A is integrated with the first section of the drift tube. Symbols S$_1$ and S$_2$ denote the central and edge drift-tube sections, respectively. 
}
\end{figure}

\section{Experimental setup}

The experimental setup realized at JINR includes a modified MaMFIS and an X-ray detector for spectroscopy studies of highly charged ions (see Figs.~\ref{fig1} and \ref{fig2}). The Dubna MaMFIS operates with stable and time-continuous electron beams characterized by the current $I_e$ in the range of  up to $50$~mA (direct current mode) and the extended electron energy $E_e$ of up to $40$~keV. The focusing system  utilizes  permanent magnets. The drift tube consists of three sections with a total length of  $30$~mm.  The typical basic vacuum is at  the level of about  $10^{-9}$~mbar.

In order to estimate the charge-state distributions of  trapped ions, the characteristic emission was recorded in the high-energy range of radiative recombination. A typical X-ray spectrum exhibits  radiation peaks originated from  ionization of different electron shells. Since energy of the recombination lines is given by the sum of electron beam energy $E_e$ and ionization energy (binding energy), the charge states of  recombinated ions can be unambiguously determined.

The uninterrupted operation of the installation in the range of $20 \leq I_e \leq 50$~mA and $8 \leq E_e \leq 22.5$~keV exceeded three months. The emission spectra were registered for 1 to 24 hours by employing a standard detector (Si) XR-100CR AMPTEK with  resolution of about $150$~eV (full width at half maximum) and the detection efficiency not exceeding $30\%$. The detector is installed at a distance of about $40$~mm from ion trap. This provides for the observation of about  $10$~mm of the electron-beam length.

\begin{figure}[t!]   
\centering
\begin{minipage}[t]{\textwidth}
\centering\includegraphics[width=0.45\columnwidth,keepaspectratio,trim=100  50  220  90,angle=0,clip]{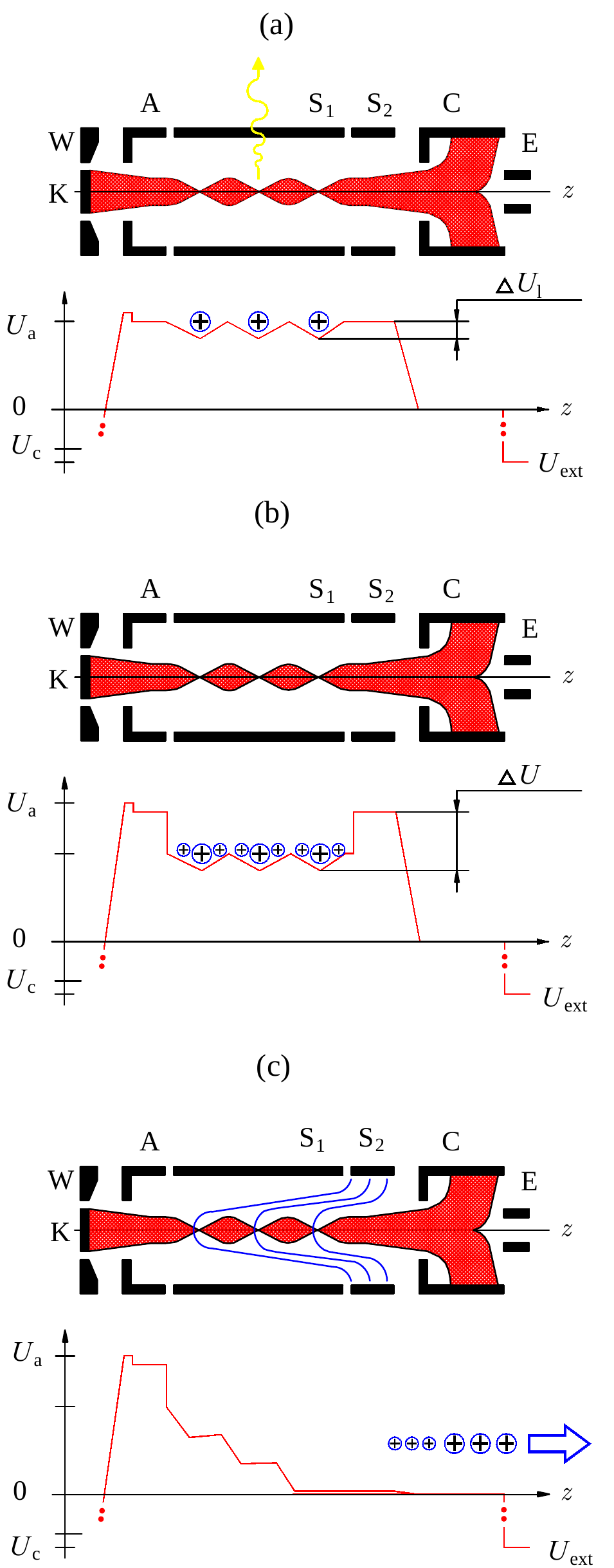}
\end{minipage}
\caption{\label{fig3} Schemes of electric potential distributions for different running modes.   (a) MaMFIS running mode. (b) and (c) Combined MaMFIS/EBIS running mode. The notations are the following: cathode K, focusing (Wehnelt) electrode W, anode A integrated with the first section of the drift tube, separate sections (S$_1$ and S$_2$) of the drift tube, electron collector C, extractor E, cathode potential $U_{\mathrm{c}}$, anode potential $U_{\mathrm{a}}$,  extractor potential $U_{\mathrm{ext}}$, depth $\Delta U_{\mathrm{l}}$ of a local trap, and depth $\Delta U$ of the common trap  $(\Delta U\geqslant \Delta U_{\mathrm{l}})$. The encircled plus signs denote highly charged ions. The local ion traps appear in the crossovers of  rippled electron beam, which are formed in focuses of  thick magnetic lens.}
\end{figure}

Besides testing the ionization ability by X-ray spectroscopy (in the running mode  indicated in Fig.~\ref{fig3}(a)), we controlled  the behavior of ions in the MAMFIS by EBIS technology (in the running modes indicated in Figs.~\ref{fig3}(b) and \ref{fig3}(c)). This can be achieved due to the fact that, for a certain length-to-diameter ratio of the drift tube, the extracting potential penetrates into the drift tube and opens local ion traps (see Fig.~\ref{fig3}(c)). Accordingly, three types of potential distributions of the electric field in the direction of the electron beam ($z$ axis) were employed in our experimental tests of the device. An equal potential applied  to all the parts of the electron-optical system (the anode and the two sections of the drift tube) provides the necessary   condition for axial confinement and subsequent ionization of highly charged ions in local traps. The radial confinement is due to the space charge of electron beam and the magnetic field.  In Fig.~\ref{fig3}(a), the depth of the  potential well is determined by the depth of the local ion traps formed by crossovers of  the rippled electron beam  ($\Delta U = \Delta U_{\mathrm{l}}$).   In Fig.~\ref{fig3}(b), the anode A and the outer section S$_2$ of the drift tube are positively biased with respect to the central section S$_1$ to form the axial potential barriers $(\Delta U > \Delta U_{\mathrm{l}})$. Due to the time dependence of the potential distributions in accordance with the EBIS technology \cite{donets67}, the ions pre-produced in discrete local ion traps can be easily extracted for their further use. In this case, the confinement time $\tau$ is determined as the time delay between sequential switches of the potential distributions depicted in Figs.~\ref{fig3}(b) and  \ref{fig3}(c).

\section{Ionization ability}

\begin{figure}[t!]   
\begin{minipage}[t]{\textwidth}
\centering\includegraphics[width=0.45\columnwidth,angle=0,keepaspectratio,clip]{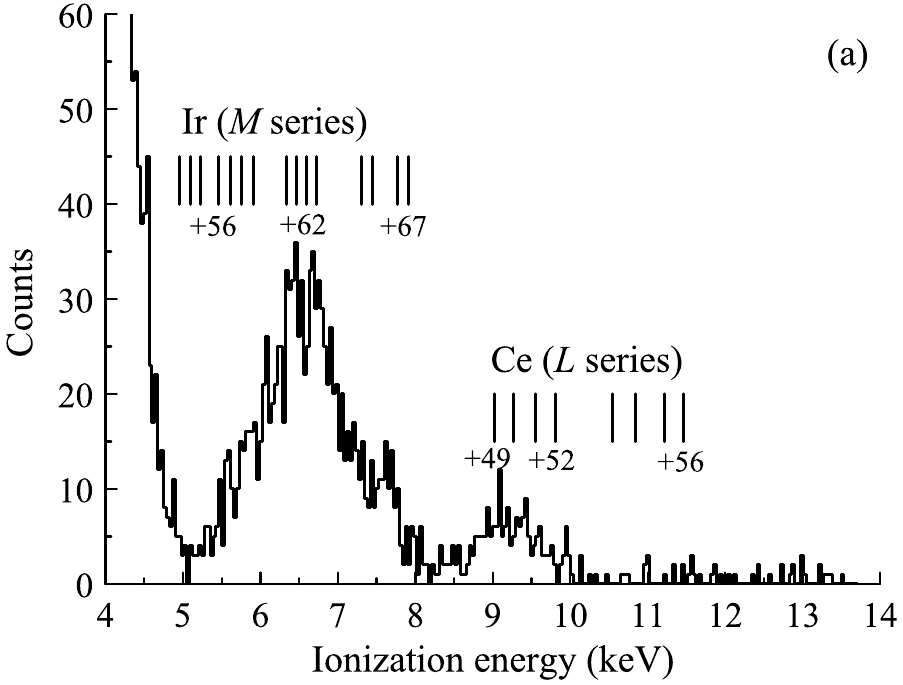}
\end{minipage}
\begin{minipage}[t]{\textwidth}
\centering\includegraphics[width=0.45\columnwidth,angle=0,keepaspectratio,clip]{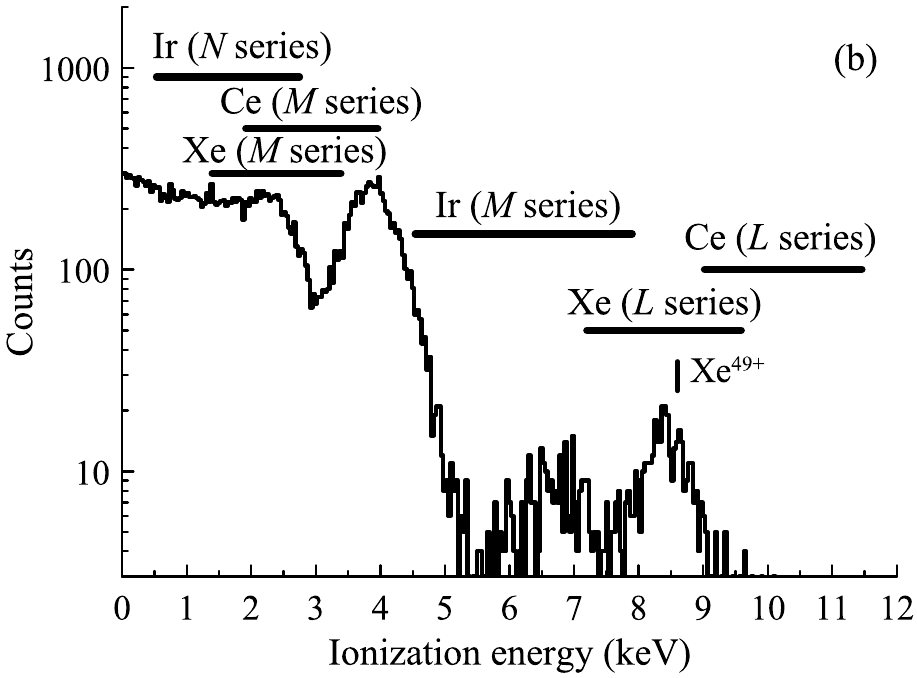}
\end{minipage}
\caption{\label{fig4} Radiative recombination spectrum for  cathode materials (a) and injected xenon (b).  The ionization energy is defined as the difference between the energy of emitted photons and the electron beam energy $E_e = 22.5$~keV.  Vacuum during the measurements was in the range from  $5 \times 10^{-9}$ to $2 \times 10^{-8}$~mbar. }
\end{figure}

In Fig.~\ref{fig4}, the X-ray emission spectra due to radiative recombination of highly charged ions of the cathode materials (iridium and cerium) and injected xenon gas with beam electrons are presented. The Dubna MaMFIS is able to ionize all atomic species up to  $M$-shell of Ir and $L$-shells of Ce and Xe. These experiments are peculiar in that they are dealing with a steady-state plasma. All  direct and reverse processes, such as  successive  ionization, charge exchange between ions and neutral atoms, radiative recombination, as well as heating of ions by electron impact and ion diffusion out of the ionization volume, are in a dynamic equilibrium. The electron current density is a key factor  affecting the ionization rates and times. A high magnitude of $j_e$ is the necessary condition for achievement of the highest charge states of  trapped ions. The maximum  ionization degrees of highly charged ions (Ir$^{67+}$, Ce$^{56+}$, and  Xe$^{52+}$) produced in the Dubna MaMFIS are consistent with the electron current density $j_e$ of the order of  $10$~kA~cm$^{-2}$. This estimate is confirmed by the computer simulations of  dominant   physical  reactions occurring during ionization of gas medium in the ion trap under  experimental conditions.  The numerical model we employed  is described in more detail in work \cite{kalagin98}.

The computer code is limited by its ability to take into consideration two ionic components only, one of which is a working substance, while the other one with a lower atomic number acts as a coolant gas. However,  in our  experiments, at least three different components are present in the ion trap. These are  hydrogen, which mainly determines composition of the basic vacuum, and two components of the cathode material (Ir and Ce), which should be also taken into account when the desired substance is injected into the MaMFIS. All the materials create a certain permanent concentration of neutral atoms in the region of the local ion trap, which cannot be controlled due to the small ionization volume. 

The dense electron beam launched by the electron gun constantly generates from the vacuum background the atomic ions with a certain charge-state distribution, starting with singly charged ions. In this case, the lower charge state ions cool  highly charged ions, a process that we termed  \emph{auto-cooling} \cite{ovsyannikov16b}.  Such process does not require a special cooling gas and does not need  the pulsed injection of the working substance. The auto-cooling was observed experimentally in work \cite{ovs10}. Obviously, this important phenomenon deserves a thorough further study  and the MaMFIS provides such an opportunity.

\begin{figure}[tbhp]    
\begin{minipage}[t]{\textwidth}
\centering\includegraphics[width=0.45\columnwidth,angle=0,keepaspectratio,clip]{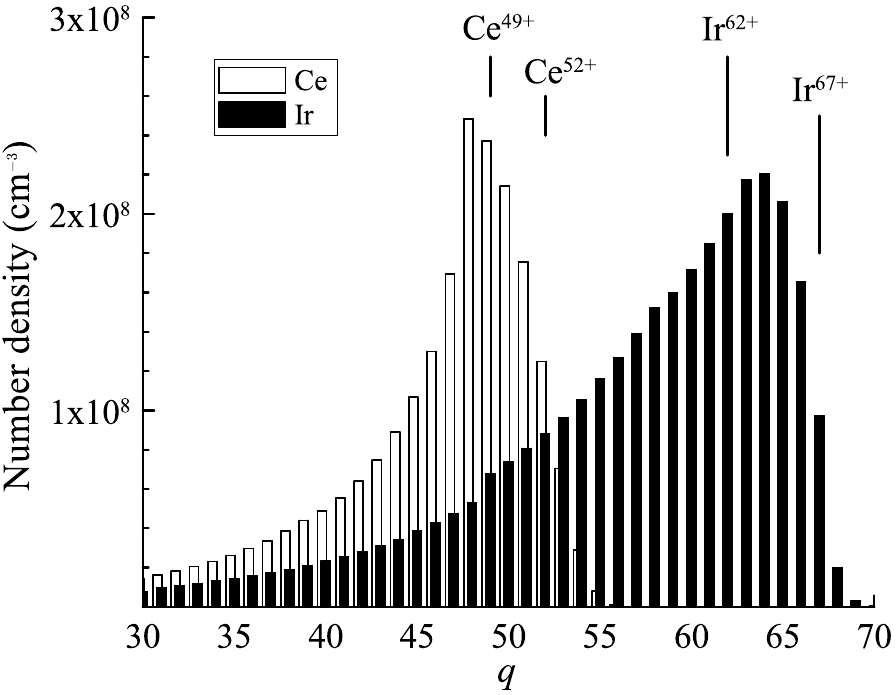}
\end{minipage}
\caption{\label{fig5}  Charge-state distributions of iridium and cerium calculated under experimental conditions upon reaching equilibrium ($E_e=22.5$~keV, $j_e=10$~kA~cm$^{-2}$, $\tau \gtrsim 150$~ms). The concentration  of residual hydrogen  corresponds to  the partial pressure  of  $5\times 10^{-9}$~mbar.  Ion charges $q$ highlighted by vertical lines are in agreement with  the experimental values obtained from X-ray emission spectrum (cf. Fig.~\ref{fig4}(a)). }
\end{figure}

The two-component model developed in \cite{kalagin98} can be employed for a fairly accurate  numerical calculations of the charge-state distributions of  trapped ions. The ion charge spectrum is determined by the dynamic balance between ionization rates, which are directly proportional to the current density $j_e$, rates of ion diffusion escape from the trap and deionization rates by charge exchange on the residual gas  and radiative recombination. For an atomic mixture of iridium, cerium and hydrogen, one can take into account the charge exchange of  iridium and cerium atoms with hydrogen only,  i.e. separately in pairwise combinations of Ir plus H and Ce plus H.  Although the collisional  processes between ions and neutral atoms of  iridium and cerium turn are neglected, it is a reasonable approximation. Indeed,  in our experiments, the concentration of hydrogen  is about two orders of magnitude higher than those of Ir and Ce atoms, which are determined by the evaporation rate of the cathode materials. In the first step, we calculated the temporal evolution of the charge-state distributions for a mixture of Ir and H, with experimental parameters of the electron beam and vacuum. When the ionization time exceeds $\sim 150$~ms, an equilibrium state is achieved with sufficient accuracy and the corresponding charge spectrum of Ir$^{q+}$ ions is determined.  In the second step, we made a  similar   procedure for a mixture of Ce and H, with the same parameters of the experiment. Finally, the charge-state distributions obtained in the first and second steps are summed up in  Fig.~\ref{fig5}.

\section{Control over ion behavior  in local ion traps}

The  electron beam current density in local ion traps can also be estimated  by studying the dependence of the radiative recombination spectra  on the ionization time. For this purpose, the running mode with a variable potential distribution, which allows one to control over the confinement time of trapped ions, as indicated in Figs.~\ref{fig3}(b) and \ref{fig3}(c), was implemented.

\begin{figure}[th!]        
\begin{minipage}[t]{\textwidth}
\centering\includegraphics[width=0.45\columnwidth,angle=0,keepaspectratio,clip]{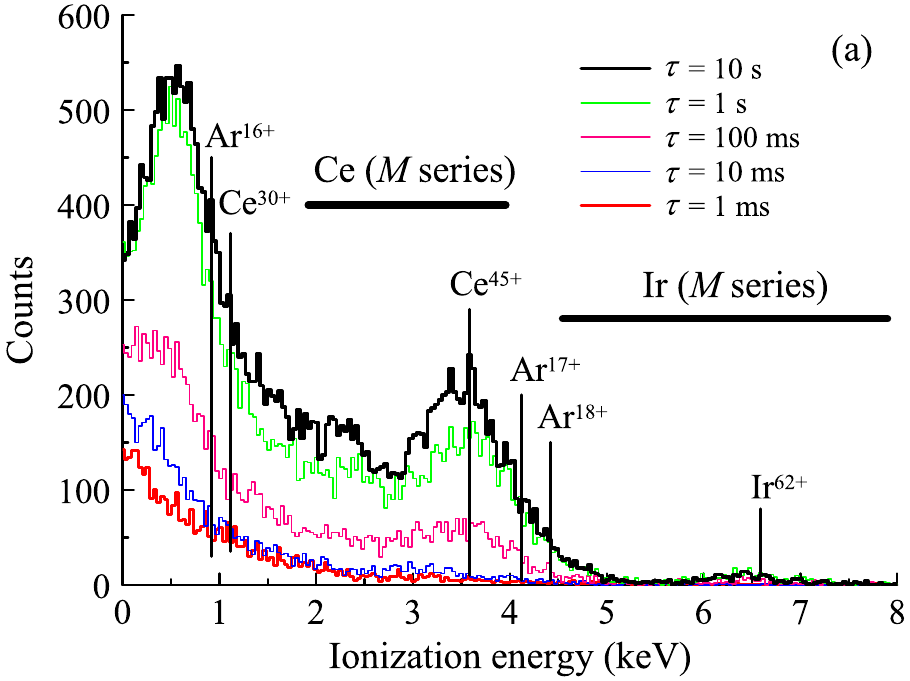}
\end{minipage}
\begin{minipage}[t]{\textwidth}
\centering\includegraphics[width=0.45\columnwidth,angle=0,keepaspectratio,clip]{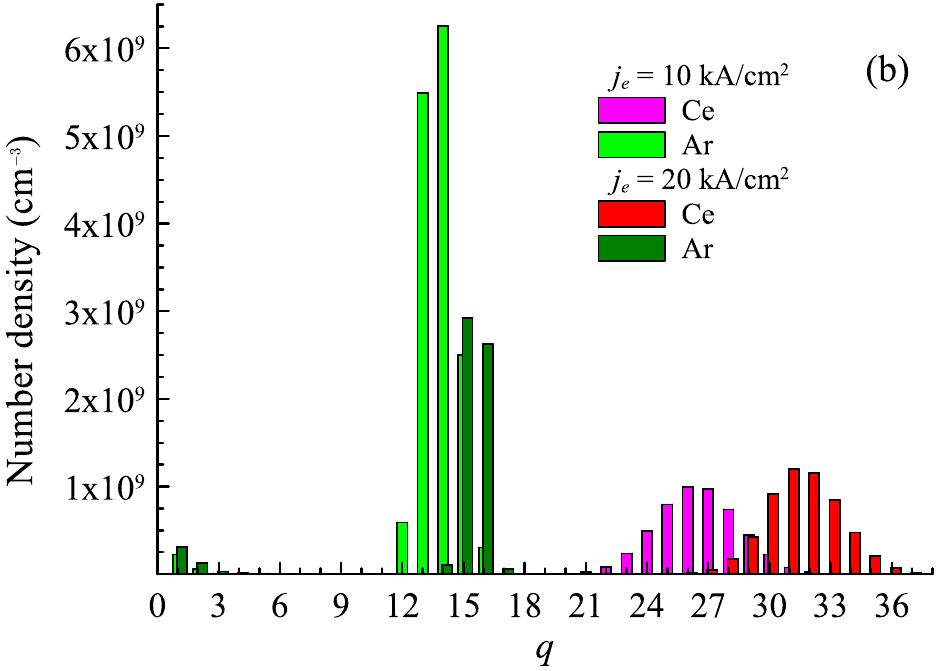}
\end{minipage}
\caption{\label{fig6}  (a) X-ray emission spectra of cathode materials and argon for different time intervals of ion trapping. (b) Computer simulations for charge-state distributions of Ar and Ce ions ($\tau =1$~ms and  $j_e  = 10$ and $20$~kA~cm$^{-2}$). }
\end{figure}

The time dependence of the  X-ray spectrum of radiative recombination for a mixture of cathode materials (Ir and Ce) and injected Ar gas is presented in Fig.~\ref{fig6}(a). The ionization degree and the number of photon counts both grow with increasing confinement time. This is in agreement with the  theoretical predictions.  An important conclusion can be drawn from the radiation spectrum for the confinement time $\tau$ of  $1$~ms.  This time  is sufficient for complete ionization of the $M$-shell of cerium and $K$-shell of argon. The measurements  can be used together with the theoretical predictions of the two-component model \cite{kalagin98} to estimate the effective electron current density $j_e$ that is experimentally achieved.  Although there are at least four different species in the trap (injected argon, pair of cathode materials, and basic hydrogen), we performed  numerical simulations of charge-state distributions for argon and cerium ions produced in 1~ms, when cerium prevails because of a higher evaporation rate.   The ion charge spectra  in Fig.~\ref{fig6}(b) are consistent with the experimental spectrum  in Fig.~\ref{fig6}(a), if  one assumes that $j_e \sim 10$~kA~cm$^{-2}$.  Note that under the same conditions a higher/lower value of $j_e$ shifts the charge-state distributions towards higher/lower values of $q$.

The control over behavior of highly charged ions in local ion traps by changing the potential distribution in the  drift tube allows one to employ the MaMFIS technology for charge breeding of low charged ions.

\begin{figure}[t!]    
\centering
\begin{minipage}[t]{\textwidth}
\centering\includegraphics[width=0.9\columnwidth,angle=0,keepaspectratio,trim=60  220  60  160,clip]{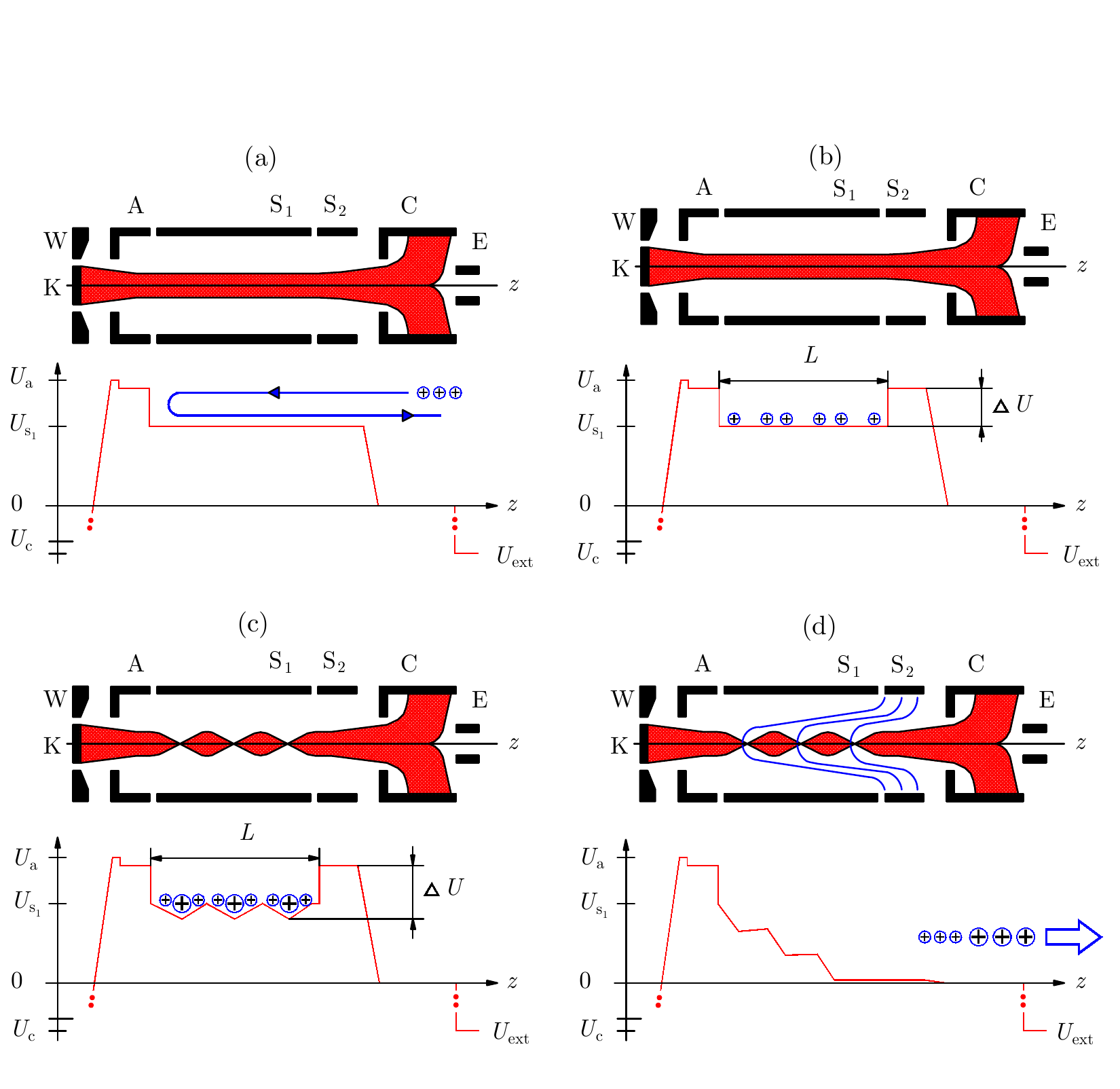}
\end{minipage}
\caption{\label{fig7}  MaMFIS as a charge breeder. Symbol $L$ denotes length of common trap, $U_{\mathrm{s_1}}$ is electric potential at section S$_{1}$ of the drift tube. Other notations are as in Fig.~\ref{fig3}.}
\end{figure}

\section{MaMFIS-based charge breeder }

In Fig.~\ref{fig7}, a new scheme for charge breeding  is presented with an indication of the electron and ion trajectories 
 and the corresponding electric potential distributions. The process of charge breeding consists of four technological stages. During the first stage,  low charge state  ions are injected from an external ion source through the extractor electrode E and the electron collector C (see Fig.~\ref{fig7}(a)). During the second stage, the low charged ions are captured into the longitudinal ion trap.  The trap is closed, when  a positive bias is superimposed on the outer section S$_2$ of  the drift tube (see Fig.~\ref{fig7}(b)). These are the same procedures as  in the standard EBIS-based charge breeder, since they are carried out in a smooth electron beam of a constant radius. In the third stage,  the smooth electron flow is transformed  into a rippled beam by varying the potential  of the focusing (Wehnelt) electrode W (see Fig.~\ref{fig7}(c)). The ions are now confined in local ion traps and  stripped of bound electrons up to the desired high charge states. This combination of the EBIS and MaMFIS technologies can make it possible to achieve an extremely high electron current density.  During the fourth stage, the highly charged ions are extracted from the breeder either in the axial  (see Fig.~\ref{fig7}(d)) or  radial direction  \cite{ovsyannikov16a}.

The efficiency of our proposed scheme for charge-state breeding requires  further experimental investigation. However, there is already experience of using a small warm electron beam ion trap as a charge breeder, in which the capture and breeding efficiencies of about $0.02\%$ for K$^{17+}$ were measured \cite{vorobjev12}. Leaving aside the problem of ion losses caused by an insufficient quality of ion optics for the injection and extraction stages,  here  we can make a few comments. 

The capacity of  the MaMFIS depends on the total length $L$ of  common ion trap (see Fig.~\ref{fig7}(b)).  The  number of  ions that can be captured into the trap per one pulse is proportional to the doubled time of flight of  an injected ion and to the ion current. With a capture time of the order of 10~$\mu$s and for the ion current as low as 1~pA, the minimum number of singly charged ions that can be trapped  is estimated to be about 60 particles.

\begin{figure}[tbh!]    
\begin{minipage}[t]{\textwidth}
\centering\includegraphics[width=0.45\columnwidth,angle=0,keepaspectratio,clip]{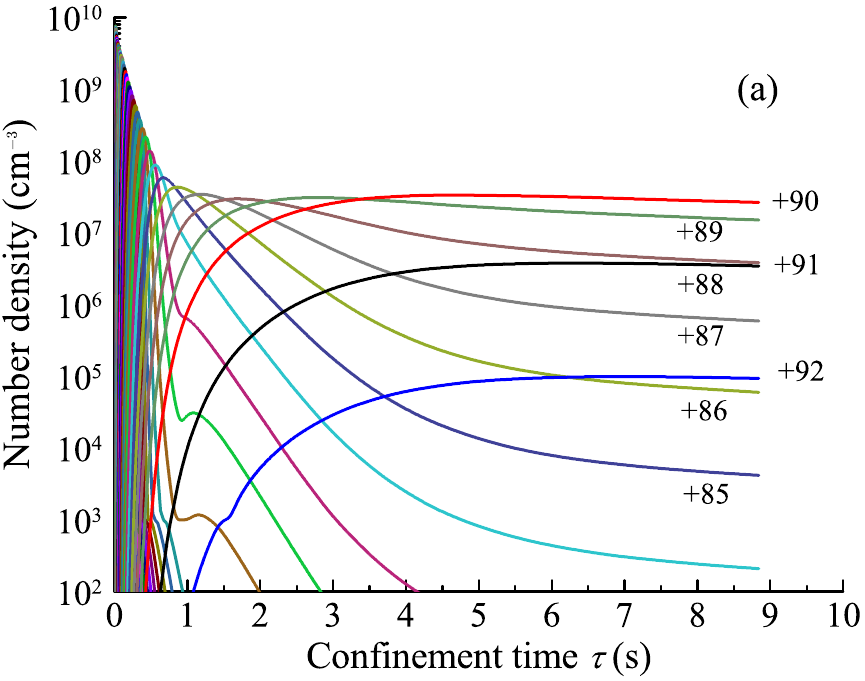}
\end{minipage}
\begin{minipage}[t]{\textwidth}
\centering\includegraphics[width=0.45\columnwidth,angle=0,keepaspectratio,clip]{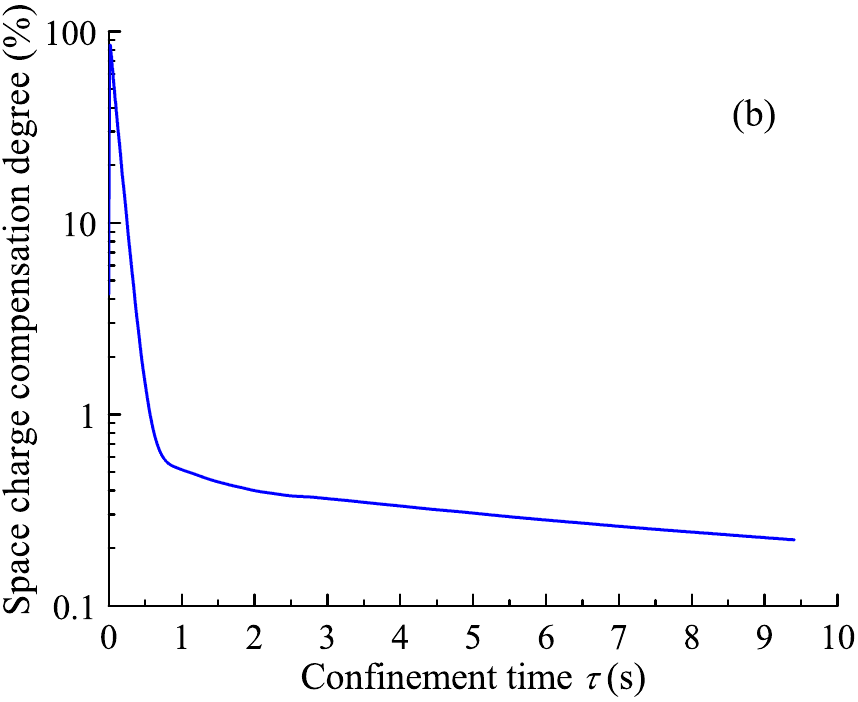}
\end{minipage}
\caption{\label{fig8}  Evolution of the charge-state distributions for U$^{q+}$ ions (a)  
and degree of the space-charge compensation for U$^{92+}$ (b)  as dependent on the confinement time $\tau$ 
($E_e =250$~keV, $I_e= 2.5$~A, $j_e = 10$~kA~cm$^{-2}$, ionization volume $\sim 2.5\times 10^{-5}$~cm$^3$, 
cooling by neutral helium at a pressure of about $10^{-10}$~mbar).}
\end{figure}

The breeding efficiency significantly depends on the particular charge states to be produced.  As a limiting case, we considered the ionization of the inner-shell electrons in uranium by using the MaMFIS technology \cite{vovs16}. In Fig.~\ref{fig8}, the time evolution of the charge-state distributions for  U$^{q+}$ ions and the degree of space-charge compensation for  bare uranium are calculated for the electron beam energy $E_e =250$~keV. As one can see, the degree of the space-charge compensation for the highest $q=+92$ declines rapidly with increasing the confinement time. This estimate for the yield of bare uranium gives 1300  nuclei per hour  \cite{vovs16}.

In contrast to the EBIS, the MaMFIS  technology, owing to its extremely high electron current density, reduces significantly the confinement time required for an efficient ionization of bound electrons. This result is of particular importance both for heavy elements and for  very short-lived radionuclides. In particular, fast ionization of inner-shell electrons can allow one to eliminate the decay channels due to  internal  conversion and electron capture (inverse beta decay), so that the half-lives of nuclides  can be increased by  orders of magnitude. Accordingly,  it seems feasible to expand the amount of short-lived radioactive species available for precision mass measurements in  Penning traps.

\section{Conclusions}

Our experiments with the upgraded MaMFIS version in Dubna confirm the high ionization ability of this technology for production of highly charged ions. The ion source has been tested at  electron beam energies of  up to $E_e = 22.5$~keV. The ionization degree  $q=+67$  of  iridium ions  was achieved. The argon atoms are ionized completely within about $1$~ms. These results are consistent with the electron current density of the order of  $10$~kA~cm$^{-2}$. Options for controlling the ion behavior in local ion traps have also been investigated. These results outline a way of how to design ion sources  with extremely short times of charge breeding, which can produce low-intensity bunches of highly charged ions at a high repetition rate.

\section*{Acknowledgments}

The authors are grateful to E.D.~Donets, E.E.~Donets, D.E.~Donets, V.V.~Karpukhin,  A.N.~Nukin,  D.O.~Ponkin, D.N.~Rassadov,  V.V.~Salnikov, and A.A.~Smirnov for their support of the experimental studies.


\end{document}